\documentclass[prx,twocolumn,aps,amsmath,showkeys,showpacs,floatfix,nofootinbib]{revtex4-1}

\usepackage{graphicx}% Include figure files
\usepackage{dcolumn}% Align table columns on decimal point
\usepackage{bm}% bold math
\usepackage{color}

\newcommand{\ket}[1]{\left|#1\right\rangle} %|"cosa">
\newcommand{\bra}[1]{\left\langle#1\right|} %<"cosa"|
\newcommand{\braket}[2]{\left< #1 \vphantom{#2} \right|
 \left. #2 \vphantom{#1} \right>} % for Dirac brackets

\def\vec#1{{\boldsymbol{#1}}}  %Vektor
\begin{document}

% ********************************************************

\title{Topological phases in small quantum Hall samples }

\author{ Tobias Gra\ss$^1$, Bruno Juli\'a-D\'iaz$^{1,2}$, Maciej Lewenstein$^{1,3}$}

\affiliation{$^1$ ICFO-Institut de Ci\`encies Fot\`oniques, 
Parc Mediterrani de la Tecnologia, 08860 Barcelona, Spain}
\affiliation{$^2$ Departament d'Estructura i Constituents de la Mat\`{e}ria,
Universitat de Barcelona, 08028 Barcelona, Spain}
\affiliation{$^3$ ICREA - Instituci\'o Catalana de Recerca i Estudis Avan\c
cats, 08010 Barcelona, Spain}
\begin{abstract}

Topological order has proven a useful concept to describe quantum 
phase transitions which are not captured by the Ginzburg-Landau 
type of symmetry-breaking order. However, lacking a local order 
parameter, topological order is hard to detect. One way to detect is via direct
observation of anyonic properties of excitations which are usually discussed in
the thermodynamic limit, but so far has not been realized in macroscopic
quantum Hall samples. Here we consider a system of few interacting bosons
subjected to the lowest Landau level by a gauge potential, and theoretically
investigate vortex excitations in order to identify topological properties of
different ground states. Our investigation demonstrates that even in
surprisingly small systems anyonic properties are able to characterize the
topological order. In addition, focusing on a system in the Laughlin state, we
study the robustness of its anyonic behavior in the presence of tunable
finite-range interactions acting as a perturbation. A clear signal of a
transition to a different state is reflected by the system's anyonic
properties.

\end{abstract}

% \pacs{67.85.De}
\maketitle

\section{ Introduction \label{intro}}

The discovery of the quantum Hall effects~\cite{klitzing,tsui,laughlin}
has ushered a new era of quantum many-body physics: Without undergoing 
the Ginzburg-Landau mechanism of spontaneous symmetry breaking, a quantum 
phase transition takes place whenever a quantum Hall system jumps from 
one to another Hall plateau upon tuning the external magnetic field 
strength. This phenomenology has been described in terms of an effective 
topological quantum field theory~\cite{wen-niu,wen}, giving rise to the 
concepts of ``topological quantum numbers'' and ``topological order'' 
which allows one to distinguish between Hall plateaus. However, this 
new form of order cannot simply be described by a local order parameter. 
In fact, it is a non-local property stemming from the highly correlated
nature of topologically ordered states, and thus robust against local
perturbations. On the one hand, this makes topological quantum matter an
interesting environment for realizing topologically protected qubits,
topologically protected quantum memory, and performing quantum
computations~\cite{nayak}. On the other hand, the measurement of topological
order is an intricate issue~\cite{alba,goldman-pnas,goldman-dauphin}.

From the theoretical point of view, progress has been made in classifying 
different types of topological matter without topological order, like 
non-interacting topological insulators of 
fermions~\cite{schnyder2008,kitaev2009}, or symmetry-protected topological 
phases of interacting bosons~\cite{wenSPT,vishwanath13}. For characterizing 
topological order, different theoretical tools have been established: 
It has first been argued that topological order manifests itself when 
the quantum state is put onto a topologically non-trivial surface 
like a torus~\cite{wen}. Depending on the genus of the surface, there 
is a different number of ways how particles can arrange themselves 
in accordance with the long-range entanglement dictated by the 
topological order. This gives rise to a characteristic number of 
degenerate ground states. In topologically trivial geometries, like 
a disk or a sphere, this classifications amounts for counting the 
number of degenerate edge states~\cite{wen,cazalilla}. In both 
cases, however, finite-range interactions and the finite size of 
the system may lift the characteristic degeneracies.

The von-Neuman entropy has been discussed as another criterion to 
characterize topological order~\cite{wen-topent,preskill-topent}. 
It has been argued that, for a topologically ordered state, it 
contains, in addition to the usual part which scales with the size 
of the system, a constant part characteristic of the topological 
order. This criterion has successfully been applied to quantum Hall
states~\cite{ee-schoutens}, but it requires the meaurement of
entropies for differently sized systems in order to extract 
the relevant information.

The most natural and direct manifestation of topological order seems 
to be the anyonic properties of the excitations, that is, the 
fractional~\cite{arovas} or even non-Abelian statistics~\cite{moore-read} 
of quasiparticle or quasihole excitations. Apart from being a signature 
of the topological order, anyons are interesting on their own, as they 
do not match with the usual classification of particles into bosons or 
fermions. Three decades after their theoretical 
prediction~\cite{laughlin,arovas}, anyons have remained rather exotic
quasiparticles: A clear experimental detection has so far only been
achieved with respect to their fractional 
charge~\cite{anyonsdetect,anyonsdetect2}, but not with respect to 
the fractional statistics. 

The difficulties in handling anyons in solid-state systems motivate 
the implementation of quantum Hall Hamiltonians in systems which 
offer a large amount of control. In the past decade, a great deal 
of attention has been paid to quantum gases. Though the atoms'
electroneutrality hinders an implementation of quantum Hall physics by real
magnetic fields, but several ways of achieving artificial magnetic fields have
been discussed and realized. Conceptually, the simplest of them is a rotation of
the atomic cloud by which a Coriolis force mimics the Lorentz force on a charged
particle in a magnetic field~\cite{schweikhard,cooper-aip,fetter-revmod}. 
Other schemes consider the generation of laser-induced Berry 
phases~\cite{dalibard,cooper12,spielmanPRL,spielman-peierls,bloch-gauge,goldrev}
, or Berry phases induced by shaking of an optical lattice~\cite{sengstock12}.
Provided with such artificial gauge fields, cold atoms should support 
different quantum Hall states, if the gauge field is sufficiently 
strong to bring all particles into the lowest Landau
level~\cite{cooperwilkin,cooper-wilkin-gunn,hafezi,brunoPRA,rapido,furukawa}. 
In particular, repulsive two-body contact interactions in a single 
component bosonic cloud ideally support the Laughlin state at 
filling factor $\nu=1/2$ (or alternatively at total angular momentum 
$L$ in $z$-direction given by $L=\hbar N(N-1)$) as an \textit{exact} 
solution. With a laser focused onto the atomic cloud, one can then 
produce Laughlin quasihole excitations~\cite{par2,bruno-njp}, and 
by adiabatically moving a quasihole created by a second laser detect 
the statistical phase of the quasiholes. As explicitly shown in 
Ref.~\cite{bruno-njp} by analysis of the Laughlin wave function, 
even for systems as small as $N\gtrsim5$ there is a bulk in the atomic 
cloud which is broad enough to pierce two quasiholes sufficiently 
far from one another and from the edge of the system, such 
that the statistical phase takes the value 1/2, expected in the 
thermodynamic limit from the plasma analogy~\cite{arovas,wen}. 
That is, at least for the case of the Laughlin state, topological 
order manifests itself even in very small systems in the form of anyons.

In this paper, we first apply this program to a larger class of 
states with topological order, in particular to the 221-Halperin state, 
ground state of two-component bosons with two-body contact repulsion, 
the $\nu=1/4$ Laughlin state, and composite fermion states~\cite{jain}. 
We find that the fractional charge of a vortex is related to the 
filling fraction of these states in the thermodynamic limit, but
accessible system sizes are too small for a clear signature of the 
fractional statistics. In the second part of this paper, we focus 
on the $\nu=1/2$ Laughlin state, as the one with the clearest anyonic 
signatures at small system size. We investigate how its anyonic 
properties are modified when the system is perturbed by some rotationally 
symmetric, finite-range interactions, keeping fixed the total 
angular momentum of the system. In this  way, our study is complementary to the
one presented in Ref.~\cite{bruno-njp}, where we investigated the role of a 
symmetry-breaking perturbation at the single-particle level. In the latter case,
the ground state is to a large extent a superposition 
of the Laughlin state plus several edge excitations, so the robustness 
of anyonic properties does not come as a surprise. 

The rotationally symmetric scenario considered here directly relates 
to the very general situation where the topological trial state is 
not an exact solution for a given interaction, as is the case in the 
``original'' fractional quantum Hall system of electrons with Coulomb 
interactions, or in the case of atoms with dipolar interactions. A 
common scenario is also the one where atoms interact with two-body 
contact interaction, but the total angular momentum is too small to 
support the Laughlin state, giving rise to large overlaps $\lesssim 1$ 
with topological states from the Read-Rezayi 
series~\cite{read-rezayi,cooper-wilkin-gunn}, or the non-Abelian spin 
singlet (NASS) series~\cite{nass-nucl,rapido,furukawa}. Usually, in 
such cases it is argued that a sizable overlap of the exact ground 
state with the trial state for a small number of particles indicates 
that the system is in the same topological phase, and this argumentation 
is often backed by considering other criteria for topological order, 
like the above-mentioned ground state degeneracy on a torus geometry, 
which is not unique unless one reaches the thermodynamic limit. In 
the example studied in this paper, we show that the topological order 
of a perturbed Laughlin state can be directly seen from the anyonic 
properties of the vortex excitations. This suggests that, despite the 
smallness of the considered system ($N=6$), a topological \textit{phase} 
is spanned within the finite Hilbert space, in contrast to a single, 
topologically ordered~\textit{state}.

\section{The system}

We consider a two-dimensional system of bosonic atoms with mass $M$, 
described by the effective Hamiltonian ${\cal H}= \sum_{i=1}^N H_i + {\cal V}$, 
where the single-particle contribution reads
\begin{align}
\label{Hi}
  H_i = 
\frac{(\vec{p}_i+\vec{A}_i)^2}{2M}  
+  
\frac{M}{2} \omega_{\rm eff}^2r_i^2 \ .
\end{align}
The kinetic term is coupled to a gauge potential $\vec{A}_i$, acting on 
the $i$th particle, and for convenience, we choose the symmetric gauge, 
$\vec{A}_i = \frac{B}{2} (y_i,-x_i,0)$, with $B$ the gauge field strength. 
Different proposals for synthesizing such gauge potential for atoms are 
reviewed in Refs.~\cite{cooper-aip,dalibard}. This first term of the
Hamiltonian thus gives rise to a Landau level (LL) structure, with 
equidistant energy levels. The second term is an effective trapping 
potential with frequency $\omega_{\rm eff}$. It can be used to control 
the $z$-component of total angular momentum of the system. For simplicity, 
we define $\hbar \omega \equiv \sqrt{\omega_{\rm eff}^2 + \frac{B^2}{4M^2}}$ 
as a unit of energy, and  $\lambda=\sqrt{\hbar/(M\omega)}$ as unit of 
length. The LL gap, $\Delta_{\rm LL}=2\eta$, and the degeneracy splitting 
of each LL due to the trap, $\delta\equiv1-\eta$, are then expressed 
in terms of the dimensionless parameter $\eta \equiv B/(2M\omega) \leq 1$.

We first consider the spinless case (bosons of only one species), and 
assume that the system can be restricted to the lowest LL (LLL). The LL 
energy spectrum is given by $E_{\ell}=\delta \ell + \rm{constant}$, 
corresponding to the Fock-Darwin (FD) states 
\begin{equation}
\phi^{\rm FD}_{\ell}(z) = {1\over \sqrt{2\pi 2^\ell \ell!}} 
z^{\ell} \exp(-|z|^2/4) \ ,
\end{equation}
with $z=x+iy$, and $\ell$ the single-particle $z-$component of the 
angular momentum. The many-body ground state in the non-interacting 
system, ${\cal V}=0$, will be a condensate wave function at zero 
angular momentum, $\ell=0$, given by
\begin{equation}
\label{cond}
\Psi_0(z_1,\dots,z_N) = \prod_{i=1}^N \phi^{\rm FD}_0(z_i) \,.
\end{equation}
Repulsive interactions may bring the system into different strongly 
correlated, topological states, depending on the range of the interactions 
and the ratio between interaction strength and effective trapping frequency. 
A realistic model for the interactions in a Bose gas is repulsive $s$-wave 
scattering, which in a many-body notation reads
\begin{align}
{\cal V}_0 = V_0 
\sum_{\ell_1,\cdots,\ell_4}
v_{\ell_1,\ell_2,\ell_3,\ell_4} a_{\ell_1}^\dagger a_{\ell_2}^\dagger a_{\ell_3} a_{\ell_4},
\end{align}
where $V_0$ parametrizes the interaction strength, and 
$a_{\ell}$ ($a_{\ell}^\dagger$) are the annihilation (creation) operators 
for a boson in state $\phi^{\rm FD}_{\ell}$. The matrix elements 
$v_{\ell_1,\ell_2,\ell_3,\ell_4}$ are given by 
$\bra{\ell_1,\ell_2} \delta(z-z')\ket{\ell_3,\ell_4} \sim \delta_{\ell_1+\ell_2,\ell_3+\ell_4}$.
Other interactions, with non-zero range, are conveniently expressed in 
terms of Haldane pseudopotentials. Therefore, one changes from a basis 
$\ket{\ell_1,\ell_2} \equiv \phi^{\rm FD}_{\ell_1}(z_1)\phi^{\rm FD}_{\ell_2}(z_2)$ 
into a basis $\ket{n,l} \equiv \phi^{\rm FD}_{n}(Z) \phi^{\rm FD}_{l}(z)$ 
with the center-of-mass coordinate $Z=(z_1+z_2)/\sqrt{2}$, and the 
relative coordinate $z=(z_1-z_2)/\sqrt{2}$. We straightforwardly find 
$\ket{\ell_1 \ell_2} =\sum_{n,l} C_{\ell_1,\ell_2}^{n,l} \ket{n,l}$ with
\begin{align}
 C_{\ell_1,\ell_2}^{n,l} =& \delta_{\ell_1+\ell_2,n+l} \sqrt{\frac{\ell_1!
\ell_2!}{n!l!}} 2^{-\frac{n+l}{2}} \times \nonumber \\ &
\sum_{k=0}{n} (-1)^{l-\ell_1+k}  \binom {n}{k}\binom{l}{\ell_1-k} \ .
\end{align}
For any isotropic interaction ${\cal V}$, we have
\begin{align}
 \bra{n,l}{\cal V} \ket{n'l'} = \delta_{n,n'} \delta_{l,l'} V_l \ ,
\end{align}
and the potential is expressed by
\begin{align}
\label{V}
 {\cal V} = & \sum_{\ell_1,\cdots,\ell_4}
\delta_{\ell_1+\ell_2,\ell_3+\ell_4} a_{\ell_1}^\dagger a_{\ell_2}^\dagger
a_{\ell_3} a_{\ell_4} 
\times \nonumber \\ &
\sum_{n,l} \delta_{\ell_1+\ell_2,n+l}
C_{\ell_1,\ell_2}^{n,l} C_{\ell_3,\ell_4}^{n,l} V_l \ .
\end{align}
This decomposition of the interaction into Haldane pseudopotentials 
$V_l$ is similar to the partial wave approach employed in scattering 
theory. Each pseudopotential $V_l$ describes the interaction of a 
pair of particles with relative angular momentum $l$. In the sum 
over $l$, for symmetry reasons, only even (odd) terms contribute 
if the particles are spinless bosons (fermions). In the case where 
the atoms have additional internal degrees of freedom, denoted by an 
index $s$, which are conserved by the interaction, the potential 
generalizes to 
\begin{align}
\label{Vspin}
 {\cal V} = & \sum_{\ell_1,\cdots,\ell_4}\sum_{s,\tilde s}
\delta_{\ell_1+\ell_2,\ell_3+\ell_4} 
a_{\ell_1s}^\dagger a_{\ell_2\tilde s}^\dagger a_{\ell_3s} a_{\ell_4\tilde s} 
\times \nonumber \\ &
\sum_{n,l} \delta_{\ell_1+\ell_2,n+l}
C_{\ell_1,\ell_2}^{n,l} C_{\ell_3,\ell_4}^{n,l} V_l^{s\tilde s}.
\end{align}
In the presence of such repulsive two-body potentials, Eqs.~(\ref{V}) 
and~(\ref{Vspin}), a relative angular momentum in the motion of two 
particles is favored in order to reduce interaction energy. The 
price to pay for this increase of angular momentum is an energy 
increase due to the trapping potential. Upon tuning either the trapping 
frequency or the interaction strength, the system therefore undergoes 
several phase transitions, from the condensate wave function, 
Eq.~(\ref{cond}), to strongly correlated phases. The most relevant 
trial wave functions to describe these phases will be presented in 
the next section.

\section{Topologically ordered phases and their excitations}
\label{ideal}

\subsection{Laughlin states}

The most famous trial state in the context of fractional quantum Hall 
systems is the Laughlin state. The wave function is given by
\begin{equation}
\label{laughlin}
\Psi_{\rm L}^q(z_1,\dots,z_N) \propto 
\prod_{i<j} (z_i-z_j)^q \exp(-\sum_i |z_i|^2/2) \,.
\end{equation}
The integer parameter $q$ fixes the total angular momentum of the state 
to $L_q=\frac{q}{2}N(N-1)$. This angular momentum is distributed such 
that the relative motion of each pair of particles carries $q$ units of 
angular momentum. Thus, every particle is seen by the other particles 
as a vortex with vorticity two. In the thermodynamic limit, where a 
filling factor $\nu$, defined as the ratio between the number of particles 
and the number of vortices, characterizes the state. The Laughlin states 
are found at $\nu=1/q$. Furthermore, the Laughlin states are zero-energy eigenstates of a potential as given in Eq.~(\ref{V}), 
if we set $V_l=0$ for $l \geq q$. This is due to the fact that 
there are no pairs of particles with relative angular momentum smaller 
than $q$ which could give a non-zero contribution to the interaction energy. 
If $V_l >0$ for $l<q$, no other state at smaller or the same total angular 
momentum has this property, and the Laughlin state becomes the true 
ground state for a sufficiently weak trapping potential. Thus, a pure 
$s-$wave scattering represents a parent Hamiltonian for the $\nu=1/2$ 
Laughlin state, and bosonic atoms in artificial gauge potentials 
therefore represent ideal systems for realizing this state. In Sec. \ref{perturbed}, we will also study a Hamiltonian with $s-$ and $d-$wave interactions, which is a parent Hamiltonian for the Laughlin state at filling $\nu=1/4$.

\subsection{Halperin states}

In the bosonic Laughlin wave function, all anticorrelation terms $z_i-z_j$ 
have to be squared for symmetry reasons. In a system with contact interaction,
however, this is not favorable from the energetic point of view, as squaring
the anticorrelation terms solely increases angular momentum. For particles 
which are distinguishable through an internal degree of freedom, one thus 
is able to find zero-energy eigenstates of the interaction at $L<N(N-1)$. 
For systems with two internal states, these are the so-called $(lmn)$-Halperin
states:
\begin{align}
\label{halperin}
&
\Psi(z_{1\uparrow},\cdots,z_{N_{\uparrow}\uparrow},z_{1\downarrow},\cdots,z_{N_
{ \downarrow}\downarrow}) \sim
\prod_{1\leq i<j \leq N_{\uparrow}} (z_{i\uparrow}-z_{j\uparrow})^l
\times \nonumber \\ &
\prod_{1 \leq i<j \leq N_{\downarrow}} (z_{i\downarrow-}z_{j\downarrow})^m
\prod_{\substack{1\leq i \leq N_{\uparrow} \\ 1\leq j\leq N_{\downarrow}}}
(z_{i\uparrow}-z_{j\downarrow})^n \ .
\end{align}
Here and in the following, we omit the overall Gaussian common to 
any state. For bosons, the smallest possible angular momentum is 
obtained at $l=m=2$ and $n=1$. The condition $l=m$ implies that the state is a spin singlet, and its filling factor is given by
$\nu=2/3$.

\subsection{Composite Fermion states}

A powerful picture to explain the fractional quantum Hall physics 
is given in terms of composite fermions~\cite{jain}. This theory 
interprets the term $\Phi = \prod_{i\neq j}(z_i-z_j)$ as an attachment 
of a vortex to each particle. These composite particles, by their 
exchange statistics fermions, may then form a non-interacting 
integer quantum Hall state, obtained 
simply by a Slater determinant $\Phi_{L'}$ over the $N$ lowest 
single-particle levels which sum up to a total angular momentum 
${L'}$. The total angular momentum of the state is then given 
by $L=L'+\frac{1}{2}N(N-1)$. For instance, 
if $L'=\frac{1}{2}N(N-1)$, we obtain the same total angular momentum 
as for the 1/2-Laughlin state, and in fact, the energetically
best choice is to put all composite particles into the LLL (running from
$\ell=0$ to $\ell=N-1$), so one obtains $\Phi_{L'}=\Phi$, and the full wave
function, $\Psi=\Phi\Phi_{L'=N(N-1)/2}$, becomes identical to the Laughlin
state. 

Apart from the Laughlin state, the composite fermion theory produces 
a series of states at smaller angular momentum, and most of them turn 
out relevant for describing bosonic gases in artificial gauge 
fields~\cite{cooperwilkin,jain-jolicoeur}. One simply has to choose
$L'<\frac{1}{2}N(N-1)$, which becomes possible if higher Landau levels are involved, also allowing for single-particle states with 
negative angular momentum, $\ell \leq -n_{\rm LL}$, where $n_{\rm LL}$ numbers the Landau level (starting with 0 for the LLL).
For that reason, the composite fermion wave function in general has to be re-projected into the LLL:
\begin{align}
\label{CF}
\Psi_{\rm CF}^{L}(z_1,\cdots,z_N) = {\cal P}_{\rm LLL} \ 
\Phi_{L'}(z_1,\cdots,z_N)
\Phi(z_1,\cdots,z_N).
\end{align}
For some $L$, there might be different ways to occupy the single-particle 
levels at the same energy. In these cases, one can construct competing 
composite fermion states. In the thermodynamic limit, the composite 
fermion wave functions describe states at filling factor 
$\nu=n/(n+1)$, with $n$ the number of occupied LLs. Below, 
we will take a closer look onto the composite fermion wave function 
at $L=N(N-2)$, representing the incompressible phase which 
is next to the Laughlin phase at $L=N(N-1)$. The composite 
fermion wave function for $L=N(N-2)$ is obtained by putting all but one 
composite fermion in the LLL, and one composite fermion into the $\ell=-1$ 
level of the first excited LL.

\subsection{Excitations}

The topological nature of a quantum state can be related to certain 
striking properties of its excitations. Quite generally, one 
distinguishes between low-lying (or gapless) edge excitations 
and gapped quasiparticle excitations. The edge excitations are 
further characterized by the finite amount of angular momentum 
(of the order of one) which is added to the ground state, whereas 
quasihole (quasiparticle) excitations change angular momentum by 
an amount of order $N$.

It has been argued that the form of the edge states characterizes 
the topological nature of a state~\cite{wen-intj}. For Laughlin-like 
phases, one has~\cite{caza}
\begin{align}
\label{edge}
\Psi_{\rm edge} = P(z_1,\dots,z_N) \ \Psi_{\rm L}^q(z_1,\dots,z_N),
\end{align}
where $P$ is any symmetric polynomial on the variables $\{z_i\}$. It 
is easy to check that if $\Psi_{\rm L}^q(z_1,\dots,z_N)$ is a zero-energy 
eigenstate of the potential ${\cal V}$, also 
$P(z_1,\dots,z_N) \Psi_{\rm L}^q(z_1,\dots,z_N)$ will be one. To count the
number of edge excitations at a given angular momentum difference 
$\delta L= L- L_q$, we should note that the symmetric polynomials are 
generated by $s_n \equiv \sum _i z_i^n$. Thus, for $\delta L=1$, there 
is a single edge state given by $s_1\Psi_{\rm L}$, for $\delta L=2$, 
there are two edge states given by $s_2\Psi_{\rm L}$ and $s_1^2\Psi_{\rm L}$, 
and so on. This sequence can be taken as a fingerprint of the topological 
phase of a system~\cite{wen-intj} and implies a degeneracy of the edge states
given by ${\cal P}(\delta L)$, which is the number of partitions of $\delta L$ 
(the number of ways in which up to $N$ non-negative integers add up 
to $\delta L$)~\cite{caza}. It is worth noting that long-range 
interactions generally lift the degeneracy of different edge states, 
with the possibility of mixing edge states with other gapped excitations.

\begin{figure*}[t]
\vspace{30pt}
\includegraphics[width=0.6\textwidth]{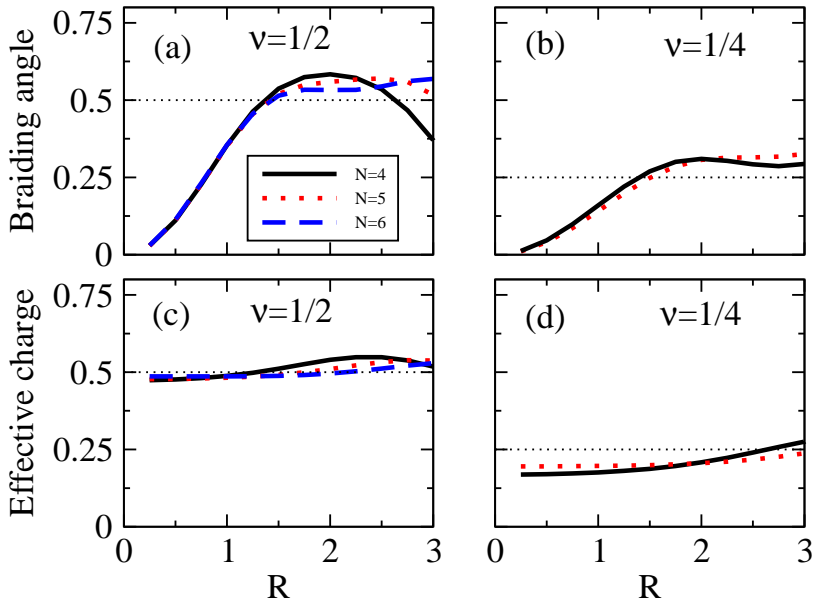}
\caption{(Color online)\label{laug2} Statistical phase angle and 
effective charge of vortices in the Laughlin states at $L=N(N-1)$ 
(a), (c) and $L=2N(N-1)$ (b), (d). $R$ is the distance between 
the hole(s) and the center of the cloud.}
\end {figure*}

In contrast to the edge excitations, there exist also localized 
excitations in form of local density increases (quasiparticles) 
or local density decreases (quasiholes) at a position $\xi$. The 
corresponding wave functions read
\begin{align}
\label{qp}
 \Psi_{\rm qp} & \sim \prod_i (\partial_i - \xi) \Psi
\equiv O_{\rm qp}(\partial_i,\xi) \Psi, \\
\label{qh}
 \Psi_{\rm qh} & \sim \prod_i (z_i-\xi) \Psi \equiv O_{\rm qh}(z_i,\xi) \Psi \,.
\end{align}
Here, we introduce ``particle'' and ``hole'' operators, $O_{\rm qp}$ and
$O_{\rm qh}$, which acting on any state create a quasiparticle or a
vortex. We note that both, $\Psi_{\rm qh}$ and $\Psi_{\rm qp}$, are states with
explicitly broken rotational symmetry, unless $\xi=0$. 

Moving $\xi$ adiabatically on a closed loop enclosing the area $A$, the wave
function acquires a Berry phase. If the system is sufficiently homogeneous, we
can write this phase factor as $\exp[i q_{\rm eff} A]$, defining an effective
charge $q_{\rm eff}$ of the quasiparticle. We will calculate this effective
charge in the following section for small-sized systems. Small dependence of
$q_{\rm eff}$ on the enclosed area suggests that the system has a topological
bulk. For topologically ordered states, a striking feature is given by the fact
that the effective charge is a fractional multiple of the elementary charge. In
case of the Laughlin state, for instance, one has $q_{\rm
eff}=e/q$~\cite{laughlin}.

Repeatedly applying the operators $O_{\rm qh/qp}$ at sufficiently 
distinct $\xi$, one obtains states with several quasihole/quasiparticle 
excitations. It then becomes possible to test the statistical 
behavior of these excitations by wrapping one around the other, 
or simply adiabatically changing the positions $\xi_1$ and $\xi_2$ 
of two excitations, such that they interchange their position. 
Again, this procedure will add a phase factor to the wave function, 
which can be written as $\exp[i q_{\rm eff} A + i \varphi_{\rm stat}]$. 
Additionally to the charge term, we now also have a second 
contribution in the exponent stemming from the statistical phase 
angle $\varphi_{\rm stat}$. This phase angle ($\rm mod \ 2\pi$) is 
0 for bosons, and $\pi$ for fermions. Quasihole or quasiparticle 
excitations of topologically ordered states are known to have 
a fractional statistical phase. Such particles are therefore 
classified as \textit{anyons}~\cite{arovas}. For the Laughlin 
states, we have $\varphi_{\rm stat} = \pi/q$.

\begin{figure*}[t]
\vspace{25pt}
\includegraphics[width=0.55\textwidth]{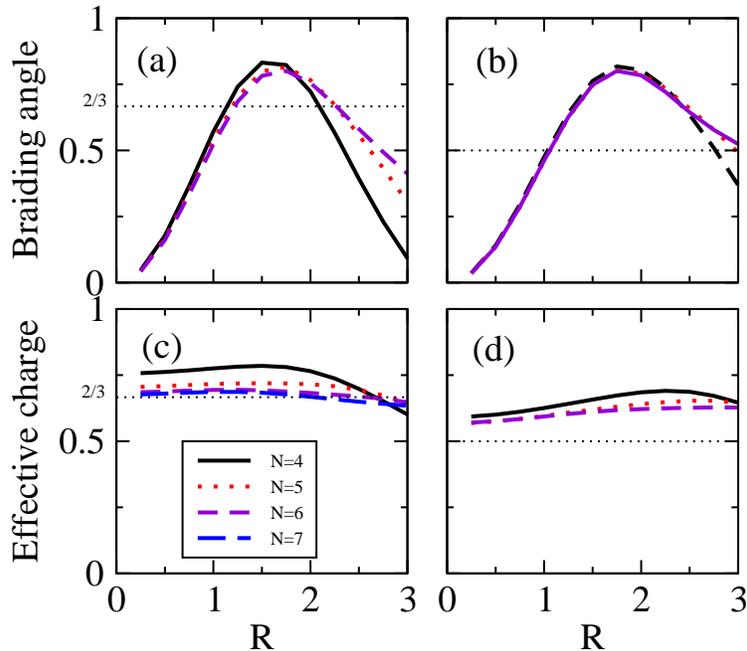}
\caption{(Color online)\label{CFfig} Statistical phase angle and 
effective charge of vortices in the $L=N(N-2)$ sector. The ones 
on the exact ground state are depicted in panels (a), and (c) while the 
corresponding ones for the Laughlin quasiparticle state 
with $\xi=0$ are depicted in (b), and (d).}
\end {figure*}

\section{Effective charge and statistical phase}
\label{calc}

\subsection{Method}

An exact analytic calculation of the effective charge and 
statistical angle of the excitations for small Laughlin 
states ($N \leq 6$) was presented in Ref.~\cite{bruno-njp}, 
in contrast with the standard literature (cf.~\cite{laughlin,arovas,wen}) 
where the effective charge and the statistical angle are 
computed in the thermodynamic limit. For small systems, one 
has to determine the function 
$f(\xi) =\bra{\Psi_{\rm Lqh}(\xi)}\nabla_\xi\ket{\Psi_{\rm Lqh}(\xi)}$, 
which in principle requires integrating all coordinates $z_i$. 
In Ref.~\cite{bruno-njp}, we avoided this multi-dimensional 
integration by decomposing the state into a many-body Fock 
basis~\cite{cpc}. Finally, one obtains the Berry phase $\gamma$ 
by simply evaluating a line integral $\gamma = \oint_A f(\xi) {\rm d}\xi$.

In this paper, we choose a slightly different approach, as we are 
interested in the behavior of quasihole excitations above generic 
states obtained by exact diagonalization. That is, we consider any 
state given in the Fock basis
\begin{align}
 \ket{\Psi} = \sum c_i \ket{i},
\end{align}
where $\ket{i} = \ket{\ell_1, \cdots, \ell_N}$ denotes the Fock states. 
To generate the quasihole state, we have to apply $O_{\rm qh}(\xi)$ to 
each Fock state. We find
\begin{align}
& O_{\rm qh}(\xi) \ket{\ell_1, \cdots, \ell_N} = \sum_{m=0}^N (-\xi)^{N-m}
2^{m/2} \times \nonumber  \\ & 
\sum_{\{p_i\}} \sqrt{\frac{(\ell_1+p_1)! \cdots
(\ell_N+p_N)!}{\ell_1! \cdots \ell_N!}} 
\ket{\ell_1+p_1, \cdots, \ell_N+p_N}.
\end{align}
Here, the sum over $p_i$ contains all choices of $p_i=0,1$ with 
$\sum_i p_i=m$.

With this, it is straightforward to calculate $\ket{\Psi_{\rm qh}}$ for 
any $\ket{\Psi}$ given in the Fock basis. For simplicity, we will assume 
that the quasiholes are moved at a fixed radial position in a rotational 
symmetric system.
The Berry phase 
$\oint_A \bra{\Psi_{\rm qh}(\xi)}\nabla_\xi\ket{\Psi_{\rm qh}(\xi)} {\rm d}\xi$ 
is then given by 
\begin{align}
\gamma =  
2\pi \lim_{\Delta\phi \rightarrow 0} 
\frac{\braket{\Psi_{\rm qh}(\phi)}{\Psi_{\rm qh}(\phi+\Delta\phi)}-1}{\Delta\phi},
\end{align}
where $\phi$ denotes the angular position of the quasihole. It is 
worth noting that for a large class of many-body states which are 
rotationally symmetric $\gamma$ solely depends on $R$ and not on $\phi$. 
Dividing $\gamma$ by $R^2$, we obtain $q_{\rm eff}$. To obtain 
$\varphi_{\rm stat}$, we apply $O_{\rm qh}$ twice at opposite positions 
$\phi$ and $\phi+\pi$, and then move both quasiholes simultaneously to 
$\phi+\Delta\phi$ and $\phi+\pi+\Delta\phi$.

\subsection{Results for the contact potential}

We will now exactly calculate the anyonic properties of vortices 
in the fractional quantum Hall wave functions introduced above. 
By comparison of these results for small systems with the known 
results in the thermodynamic limit, we find that despite significant 
finite-size effects, anyonic properties and thus topological order 
are established already in systems as small as $N\approx 6$. 

\begin{figure}[t]
\vspace{10pt}
\includegraphics[width=0.4\textwidth]{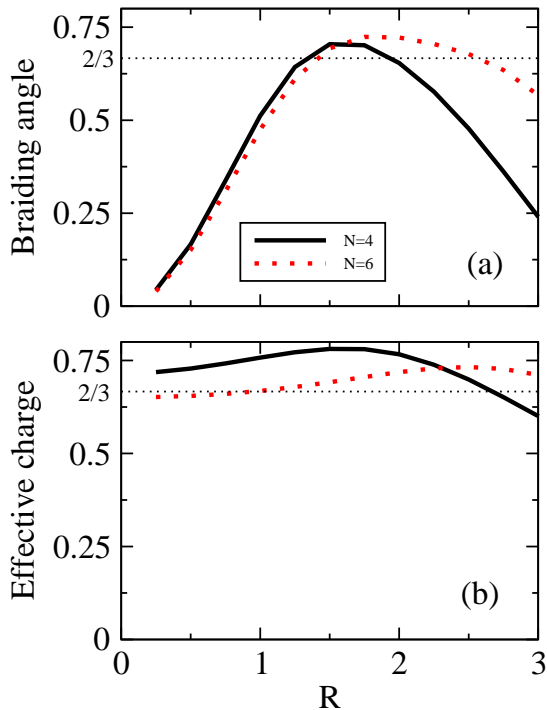}
\caption{(Color online)\label{halp} Statistical phase angle (a) 
and effective charge of vortices (b) in the $(221)$-Halperin state.}
\end {figure}

\subsubsection{Laughlin states}

The exact ground state of a single component bosonic system with contact 
$s-$wave atom-atom interactions at $L=N(N-1)$ is the Laughlin state, 
Eq.~(\ref{laughlin}), for $q=2$. As described above, we introduced up 
to two vortices into the ground state wave function, and calculated 
the effective charge and braiding angle for different system sizes,
$4\leq N \leq 6$. The results are shown in Fig.~\ref{laug2}(a,c): A 
single vortex at any radial position $R$ sufficiently far from the 
edge has an effective charge which is close to $1/2$, the value expected 
for the thermodynamically large system. Naturally, the effective 
charge drops when the particle density diminishes at the edge of 
the system. The same is true for the braiding angle of two quasiholes, 
but here one also has to assure that they are sufficiently far 
from each other. Two vortices close to the center overlap with each 
other, distorting the value for the statistical phase. This makes 
it difficult to extract the bulk behavior for very small systems. 
The braiding angle for $N=4$ is basically a monotonously growing 
function of $R$, until a value close to $1/2$ is reached and the 
function starts to monotonously decrease again. However, by comparing 
the curves for different number of particles, one clearly finds that a plateau is formed near $1/2$, that is, a sizable bulk with the 
expected anyonic property is formed.

As shown in Fig.~\ref{laug2}(b,d), the properties of vortices in 
small-sized $q=4$ Laughlin states differ more strongly from the 
thermodynamically expected values of $1/4$. Nevertheless, comparison 
of the figures for 4 and 5 particles shows that the numbers quickly evolve 
in the expected direction. Due to the large angular momentum 
of this state, the Hilbert space grows too fast with $N$ to consider 
larger systems. On the other hand, even for $N=5$, the non-monotonic 
behavior of the statistical phase around 0.3 suggests that an 
extended bulk has developed.

\subsubsection{Composite fermion state vs. Laughlin quasiparticle 
state at $L=N(N-2)$}

If one increases the trap frequency (or decreases the interaction 
strength), at some critical value the system undergoes a transition 
from the Laughlin state to a state where the total angular momentum 
is decreased by $N$ units to 
$L=N(N-2)$~\cite{cooperwilkin,brunoPRA,bruno-njp}. At this angular 
momentum, one can construct a composite fermion state, Eq.~(\ref{CF}), 
by putting all but one composite particles into the lowest Landau level. 
Accordingly, this composite fermion state describes a system at filling 
factor $\nu=2/3$, and the vortex excitation (despite not being the 
elementary excitations of this state) should carry this effective charge 
and fractional statistics~\cite{jain-book}. It has been shown that the 
exact ground state has very large overlaps with this composite fermion 
state ($\gtrsim 0.99$) for a small number of particles~\cite{cooperwilkin}. 
On the other hand, at $L=N(N-2)$ one can also construct a quasiparticle 
excitation of the Laughlin state according to Eq.~(\ref{qp}), with the 
quasiparticle at the center, $\xi=0$. Considering system sizes of $4\leq
N\leq 7$ particles, we also find good overlaps ($\gtrsim 0.97$) between 
this quasiparticle state and the true ground state. A vortex in the Laughlin
quasiparticle state, however, should be of the Laughlin-type, that is, 
showing $1/2$-type anyonic properties.

Thus we encounter a situation where simply from overlap arguments 
it is not yet clear which topological properties can be expected, 
as two trial states describing different topological phases compete 
with each other. It is obvious that both, composite fermion and 
Laughlin quasiparticle state also have very large overlaps one 
with another. One could expect that in such a situation the topological 
properties are not well defined unless we go to larger systems. 
However, at least with respect to the fractional charge of a vortex 
in the true ground state, shown in Fig.~\ref{CFfig}(a,c), the result 
of our direct diagonalization of the Hamiltonian approaches $2/3$ when
increasing the number of particles from 4 to 7. 
On the other hand, the braiding angle does not yet show a clear 
bulk behavior for $N=6$.

It is interesting to compare these results with the ones shown in 
Fig.~\ref{CFfig}(b,d), obtained for the Laughlin quasiparticle 
wave function. Despite the sizable overlap with the true ground 
state, the vortex charge is shifted towards a significantly lower 
value around 0.57 for $N=7$, and thus closer to the expected value 
$1/2$ than to the composite fermion value $2/3$. As the Laughlin 
quasiparticle state has, compared to the Laughlin state, a locally 
increased density around the center, it is not surprising that also 
the effective charge of a nearby vortex is increased.

\subsubsection{$(221)$-Halperin state}

To conclude this section, let us finally present the charge and 
statistics of vortices in the $(221)$-Halperin state, Eq.~(\ref{halperin}). 
For this state, which is the spin-singlet generalization of the 
$1/2$-Laughlin state, the elementary excitations are given in terms 
of vortices (and quasiparticles). As the $(221)$-Halperin state 
describes a system at $\nu=2/3$, we expect the fractional properties 
of the vortices to be given by that number. In fact, as shown in 
Fig.~\ref{halp}, for $N=6$, the effective charge has converged towards 
this value, and the braiding angle function is about to exhibit a plateau 
slightly above $2/3$.

\section{Perturbing the bosonic 1/2-Laughlin state\label{perturbed}}

\begin{figure}[t]
\includegraphics[width=0.44\textwidth]{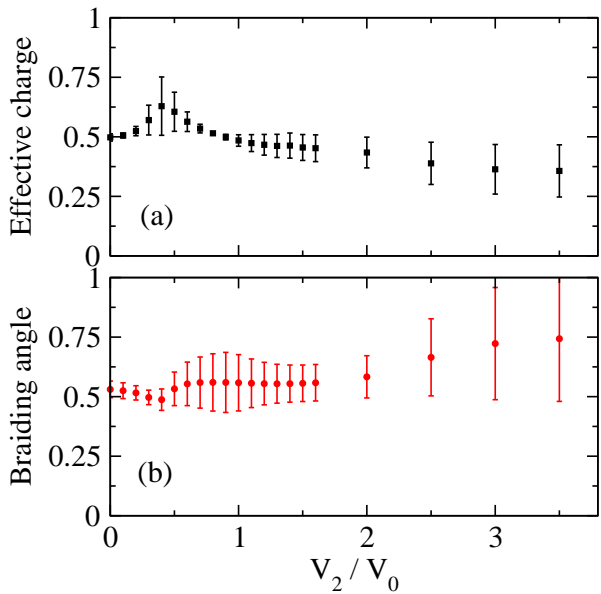}
\caption{(Color online)\label{charge_angle} Effective charge (a) and 
statistical phase angle (b) of vortices averaged over different 
vortex positions. The represented value is the average $\pm$ the 
standard deviation, in a $\nu=\frac{1}{2}$ Laughlin system 
($N=6$, $L=30$) perturbed by finite-range interactions 
characterized by a Haldane pseudopotential with relative strength $V_2/V_0$.}
\end{figure}

In the previous section, we have analyzed different fractional quantum 
Hall states which are obtained as ground states of a two-body contact 
interaction in a Hilbert space spanned by the lowest Landau level at 
a given $L$. For $L=N(N-1)$, one obtains the 1/2-Laughlin state as 
the unique zero-energy eigenstate with clear signatures of anyonic 
order even for small $N$. In this section, we analyze the robustness of 
these properties against deformations.

To this aim, we will study a Hamiltonian which differs from the parent
Hamiltonian ${\cal H}_{\rm parent} = \sum_i H_i + {\cal V}_0$ by some
controllable perturbation. With this, the ground state of the Hamiltonian 
will, quite generally, read $\Psi_{\rm GS} = 
\alpha \Psi_{\rm L} + \beta \Psi_{\rm corr}$, with $\alpha$ and 
$\beta$ non-zero parameters depending on the strength of the perturbation.

In Ref.~\cite{brunoPRA}, a similar study has been performed, motivated 
by the fact that a laser-induced gauge field for atoms is equivalent
to a real magnetic field only after an adiabatic approximation. The
non-adiabatic effects can then be taken as a perturbation to the parent
Hamiltonian, breaking the rotational symmetry. The ground state is then 
a superposition of polynomials at different $L$, and in Ref.~\cite{brunoPRA} 
it has been approximated by a wave function $\Psi_{\rm L} (\alpha + \beta \sum
z_i^2)$. Thus, in that case, the correction to the Laughlin state is 
given by an edge excitation of the Laughlin state, cf. Eq.~(\ref{edge}). 
From this it is directly clear that the bulk physics remains unchanged, 
and quasiholes which are sufficiently far from the edge maintain the 
Laughlin-like behavior. In this sense, the system considered in 
Ref.~\cite{brunoPRA} trivially realizes an extended Laughlin \textit{phase}.

The situation which we wish to study now is less trivial, as we will 
assume a perturbation which does \textit{not} break any symmetry. 
Thus, the ground state $\Psi_{\rm GS}$ is obtained within the same 
Hilbert space as the Laughlin state $\Psi_{\rm L}$. Can the anyonic 
properties be maintained in such a scenario? We will study this question 
by adding a finite-range interaction, in terms of a non-zero Haldane 
pseudopotential ${\cal V}_2$ to the Hamiltonian. Such a model interaction 
approximates well a combination of a contact potential and a rapidly 
decaying long-range potential like $r^{-3}$ dipolar interactions. By 
changing the strength of one potential relatively to the other, the 
ratio $V_2/V_0$ of the two pseudopotential strengths can be tuned. In 
particular, by making the $s$-wave scattering attractive, it is also 
possible to make $V_2\geq V_0 \geq 0$. A study on the 
torus~\cite{cooper-dipolar} has shown that a symmetry-breaking phase 
transition at $V_2/V_0 \approx 0.5$ brings such system from a 
Laughlin-like phase to different vortex lattice phases. In this 
paper, we exclude the option of symmetry breaking by fixing the total 
angular momentum to $N(N-1)$, that is, regardless of the possibility 
to lower energy by increasing angular momentum, we study how a Laughlin 
system evolves when finite-range interaction are turned on.

\begin{figure}[t]
\includegraphics[width=0.5\textwidth]{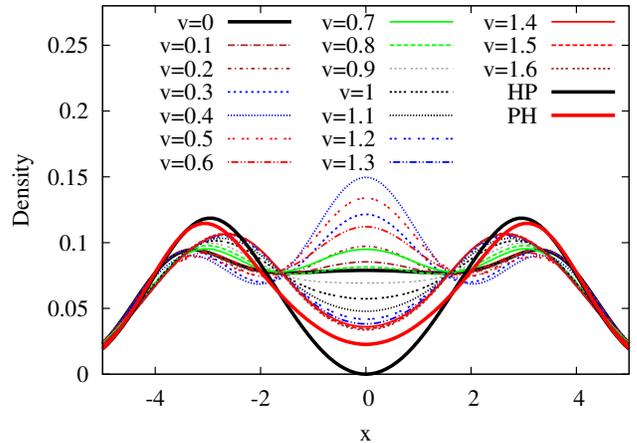}
\caption{(Color online)\label{densprof} Density profiles of the 
ground state of the system ($N=6$, $L=30$) for different Haldane 
pseudopotential strengths $v=V_2/V_0$. The curves denoted by HP 
and PH show the density profiles of a trial wave function obtained 
from the Laughlin by adding a quasiparticle and subsequently a 
quasihole (PH) or a quasihole and subsequently a quasiparticle 
(HP) in the origin.}
\end{figure}

\begin {figure}
\vspace{20pt}
\includegraphics[width=0.44\textwidth]{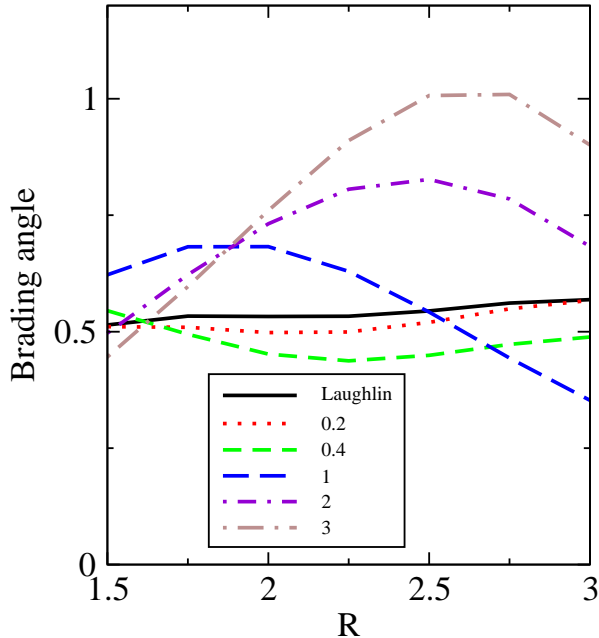}
\caption{(Color online)\label{braiding} Statistical phase ($N=6$,
$L=30$) as a function of radial vortex position $R$ for different 
Haldane pseudopotential strengths. The labels correspond to the 
values of $V_2/V_0$.}
\end {figure}

\subsection{Braiding of excitations}

In order to get a quick overview of the quasihole properties when 
finite-range interactions are switched on, we plot for $N=6$ in 
Fig.~\ref{charge_angle} average values of the effective charge and 
the braiding angle as a function of $V_2/V_0$, where the average 
is about the radial position of the quasiholes. We consider the 
interval $R \in [0.25,3]$ for the charge, and the interval 
$R \in [1.25,3]$ for the statistical phase, in steps of 
$\Delta R= 0.25$. In particular, we display an 
errorbar for each point in the figure given by its standard deviation in the interval. 
In this way we can discern between genuine fractional behavior, 
which has well-defined bulk properties and thus a small standard deviation, and other regimes. For 
instance, as we saw in Fig.~\ref{laug2}, for the Laughlin state 
of $N=6$, the fractional charge and statistical phase are 
mostly constant in a broad region of $R$. This is reflected 
in a very small standard deviation of the charge and braiding 
angle at $V_2/V_0=0$ in Fig.~\ref{charge_angle}(a,b). 

For small values of $V_2/V_0<0.4$ the situation is very similar to 
the pure Laughlin state, with robust fractional properties of the 
ground state. At $V_2/V_0 \approx 0.4$, the situation changes, and the 
average effective charge reaches its global maximum at 
$q_{\rm eff} \approx 0.64$, that is, almost 30\% off the Laughlin 
value, and at the same time also its standard deviation increases 
dramatically. For larger values of $V_2/V_0$, the effective charge 
decreases monotonically, and also its the standard deviation goes 
back to a very small value (around $V_2/V_0 \approx 0.8$) before 
then increasing again. This behavior can be traced to density 
changes of the system upon increasing $V_2/V_0$, shown in 
Fig.~\ref{densprof}. While the Laughlin system, characterized by a 
homogeneous bulk density, first develops a density peak in the center, 
when $V_2/V_0$ is increased to 0.4, this peak then continuously 
shrinks down to a density minimum for larger $V_2/V_0$. During this 
shrinking process, the system goes through a state of homogeneous 
density around $V_2/V_0 \approx 0.8$.

The braiding properties turn out to be Laughlin-like up to a value
$V_2/V_0 \approx 0.4$, where the averaged braiding angle has a global 
minimum around $V_2/V_0=0.49$. The standard deviation still remains small, in contrast to the large standard 
deviation of the fractional charge. This indicates that 
the braiding angle would be a more robust property to characterize
the topological properties of the system in this case. For larger 
values of $V_2/V_0$ the braiding angle increases slightly, but remains 
close to 1/2 up to $V_2/V_0 \approx 1.5$. One has to note, however, 
that for any $V_2/V_0 \gtrsim 0.5$, the standard deviation of the 
braiding angle is significantly higher than in the Laughlin regime. 
To illustrate this point, we have explicitly plotted in 
Fig.~\ref{braiding} the braiding angle as a function of the vortex 
position for different values of $V_2/V_0$. For $V_2/V_0<0.4$ the 
curves are Laughlin-like, with non-monotonic behavior around an average value 
close to 0.5, indicating the formation of a bulk. The curves  
for $V_2/V_0 \simeq 1$ despite having an average value close to 1/2, 
do not show any bulk behavior. They monotonically increase to some 
maximum value significantly above 1/2, and then monotonically decrease 
until reaching the edge of the system. This behavior results in an 
increased standard deviation. For large $V_2/V_0$ the curves further increase, leading to average values above the Laughlin mean, from 0.55 at 
$V_2/V_0=1.5$ to 0.75 at $V_2/V_0=3.5$.

Summarizing, Laughlin-like behavior with respect to both fractional 
charge and statistics is found only up to some value $V_2/V_0$ below 
0.4, whereas the bulk behavior with respect to quasihole braiding 
is lost only for slightly higher values of $V_2/V_0$. The analysis of 
the excitation thus yields $V_2/V_0=0.4$ as a rough estimate of the 
topological phase transition. It suggests that there is no topological 
order in the system when the Laughlin order has been destroyed, but 
the size of the considered system ($N=6$) might be too small to display 
the order. To back our interpretation so far, and get further insight 
into the physical behavior of the system, we will in the following discuss the 
overlaps, correlation functions, and the energy spectrum of the system.

\subsection{Overlaps, correlations, and spectral properties}

\subsubsection{Low lying spectral properties}

\begin {figure}[t]
\includegraphics[width=0.49\textwidth]{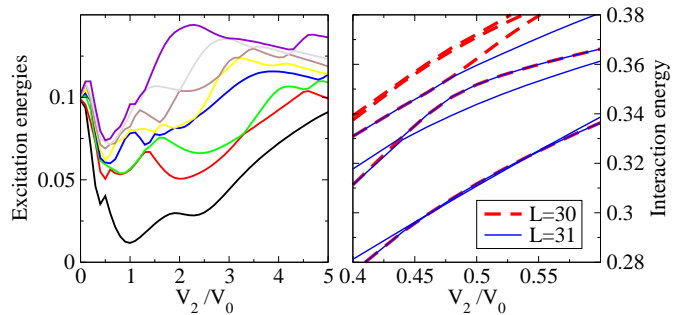}
\caption{(Color online)\label{spectra} (a) Gaps above 
ground state at $N=6$ and $L=30$ in the presence of a Haldane 
pseudopotential with relative strength $V_2/V_0$. (b)
Energies of lowest states at $N=6$ and $L=30$ and $L=31$.}
\end {figure}

A first quantity worth examining is the energy spectrum. We 
distinguish between the spectrum at constant angular momentum 
$L=N(N-1)$, and the edge spectrum in subspaces at higher $L$. 
We will increase the ratio $V_2/V_0$ keeping $V_0$ constant 
and increasing $V_2$. It is thus natural that we find that the 
energy of all levels increases with $V_2/V_0$. In contrast to this, 
the energy gap above the ground state in the $L=N(N-1)$ subspace 
turns out to be larger at $V_2=0$. Remarkably, as shown in 
Fig.~\ref{spectra}(a), the gap never closes in the interval 
$0 \leq V_2/V_0 \leq 5$. Thus, we do not obtain a clear hint 
for a phase transition in the form of a level crossing. The only 
signatures of some reorganization are the three minima of the 
gap. They can be traced back to avoided level crossings: The 
global minimum, at $V_2/V_0 \approx 1$, coincides with an 
avoided level crossing between the ground state and its 
first excited state. Near the other minima, at $V_2/V_0 \approx 0.4$ 
and $V_2/V_0 \approx 2.4$, avoided level crossings between the 
second and the third level are pronounced, whereas the ground 
state level remains relatively far from the neighboring level.

While these observations do not suggest a phase transition, a 
different picture emerges from the edge spectrum, that is, the 
spectrum of $L$-changing excitations. Here, we restrict ourselves 
to $\Delta L =1$. For almost the full range  $0 \leq V_2/V_0 \leq 5$, 
we find the ground state energy at $L=N(N-1)$ to be the same as 
at $L=N(N-1)+1$, and the ground state at $L=N(N-1)+1$ is obtained 
from the ground state at $L=N(N-1)$ as an edge state via the 
construction of Eq.~(\ref{edge}). The gap above the ground state 
is slightly smaller in the $L=N(N-1)+1$ subspace, but of similar 
order in both subspaces. There is, however, one remarkable exception 
to this behavior, found in the interval $0.4 \leq V_2/V_0 \leq 0.6$, 
depicted in Fig.~\ref{spectra}(b). Here, the gap in the $L=N(N-1)+1$ 
subspace vanishes completely. Two nearby level crossings between the 
two lowest states even result in a small negative gap between the 
edge state and a second state. Since in quantum Hall effect the 
robust conductivity properties characterizing the different phases are 
attributed to the number of robust edge states, the occurrence of a 
second low-lying excitation strongly suggests a topological phase 
transition located in the interval $0.4 \leq V_2/V_0 \leq 0.6$, in 
agreement with our considerations of the anyonic properties.

\subsubsection{Overlaps with perturbed Laughlin wave functions}

\begin {figure}[t]
\includegraphics[width=0.5\textwidth]{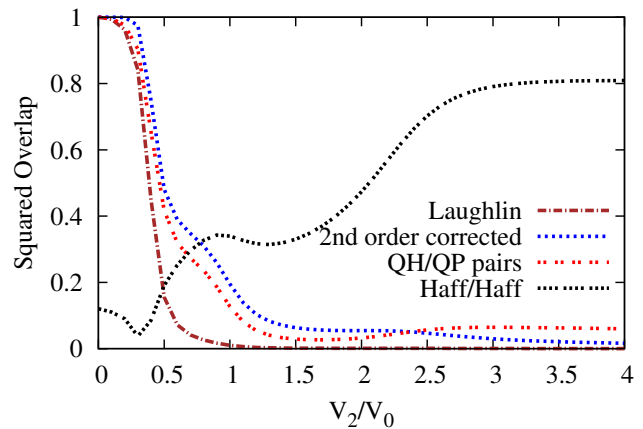}
\caption{(Color online)\label{overlaps} Overlaps of the exact ground state at
$N=6$ and $L=30$ with different trial wave functions in the presence of a
Haldane pseudopotential with relative strength $V_2/V_0$.}
\end {figure}

Next, we calculate the overlap of the exact ground state (for $N=6$) 
with the Laughlin state, shown in Fig.~\ref{overlaps}: It monotonously 
decreases with increasing $V_2/V_0$, but it remains above $>0.9$ up to 
values $V_2/V_0 \approx 0.3$. For stronger $d$-wave interactions, the 
decay becomes significantly steeper. The sharpest decline is found in 
the interval $0.4<V_2/V_0<0.5$, in which the overlap falls from 0.672 
to 0.395. This observation suggest a topological transition at 
$V_2/V_0 \approx 0.4$, backing our estimate of the phase boundary 
obtained by analysis of the quasihole behavior.

\begin {figure*}[t]
\includegraphics[width=0.6\textwidth, angle=-90]{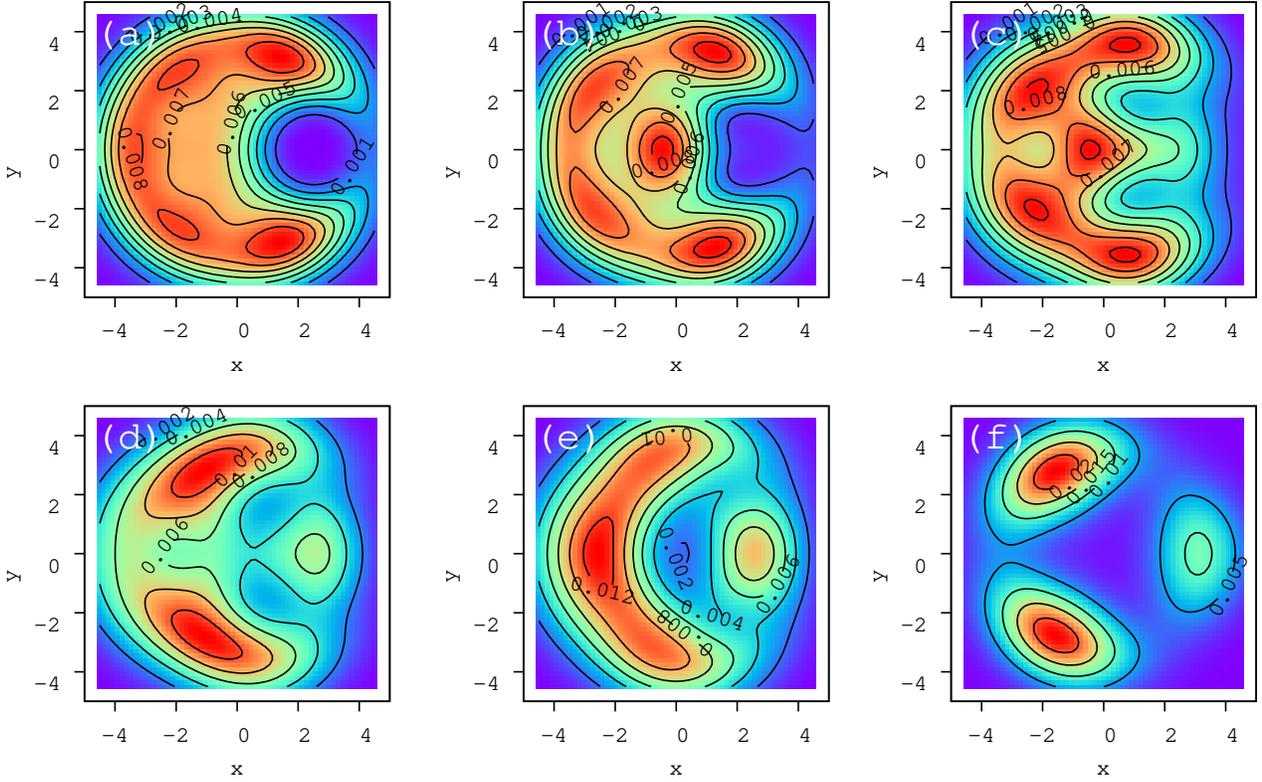}
\caption{(Color online)\label{corrs} Ground state correlation functions 
for $N=6$ at different values of $V_2/V_0$ with one particle fixed 
to $(x,y)=(2.5,0)$. Bright colors denote relatively large values of 
the correlation function, that is, a high probability of finding a 
particle at the respective position. From (a) to (f), $V_2/V_0$ takes 
the values: 0, 0.3, 0.5, 0.8, 1.5, 4.0.}
\end {figure*}

The absence of a level crossing with respect to the ground state 
suggests the use of perturbation theory to describe the ground
state. Making use of the numerical solutions at $V_2/V_0=0$, we may 
try to describe the ground state at finite $V_2/V_0$ as a perturbative 
state. As long as this ansatz yields good overlaps with the exact 
ground state, the system can be interpreted as a perturbed Laughlin 
system. As we also have exact knowledge of the ground state at finite 
$V_2/V_0$, we can directly check to which order in perturbation 
theory we get improvements. It turns out that for any $V_2/V_1 > 0.1$, 
results improve up to the second order, but then become worse. The 
reason is the quickly diverging nature of the perturbative expansion: 
Setting the smallness parameter $V_2/V_0=1$, the first-order corrections 
to the ground state, $\Psi^{(1)}$, are normalized to 0.48. The
second-order corrections, $\Psi^{(2)}$, are normalized to 1.26, whereas 
the third-order corrections are normalized to 139. We thus define
\begin{align}
 \Psi_{\rm pert} \sim  \Psi_{\rm L} + \frac{V_2}{V_0} \Psi^{(1)} +
\left(\frac{V_2}{V_0}\right)^2 \Psi^{(2)},
\end{align}
and plot the overlap of $\Psi_{\rm pert}$ with the exact ground state in 
Fig.~\ref{overlaps}. Perturbation theory works very accurately 
up to $V_2/V_0 \approx 0.3$ with overlaps $\gtrsim 0.99$. The results 
at larger $V_2/V_0$ become increasingly worse, and the steepest decay 
occurs between $0.4 \leq V_2/V_0 \leq 0.5$, as has already been the case 
for the overlap of the exact ground state with the unperturbed Laughlin 
state. For $V_2/V_0=0.5$, the squared overlap has dropped below 0.5, 
that is, the dominant contribution to the ground state can no longer 
be understood as a perturbed Laughlin state.

While the perturbative ansatz yields a good description of the ground 
state for sufficiently small $V_2/V_0$, it does not tell much about the 
nature of the deformations. To get a physical picture, we construct 
different trial states. One possibility of deforming a Laughlin state 
without changing its angular momentum is the creation of one 
quasiparticle/quasihole pair in the center, yielding the trial states 
$O_{\rm qh}(0)O_{\rm qp}(0) \Psi_{\rm L}$ and $O_{\rm qp}(0)O_{\rm qh}(0) \Psi_{\rm L}$. 
Further states can be obtained by repeatedly generating quasiparticle/quasihole 
pairs at the center. The overlap of a trial state constructed as a superposition of the 
Laughlin state, the two states with one pair excitation,
and the four states with two pair excitations is shown in Fig.~\ref{overlaps}.
This construction provides a significant improvement compared to the pure
Laughlin state, but is less accurate than $\Psi_{\rm pert}$.

Why the system, in the presence of $d$-wave repulsion between the atoms, 
benefits from the creation of quasiparticle/quasihole pairs can be 
understood by noting that this construction takes out anticorrelations 
from the state via the derivative associated to the quasiparticle 
construction, and reintroduces new anticorrelations via the $z$'s 
of the quasihole construction. In this way, some pairs of particles gain relative angular momentum, and thus will avoid $d$-wave scattering, for the cost of an increased $s$-wave interaction between other pairs of particles.
Alternatively, we may achieve a similar effect in a more explicit way by writing
\begin{align}
 \Psi_{\rm trial} = 
\sum_{\rm P} {(z_{{\rm P}(1)} - z_{{\rm P}(2)})^2 \over 
(z_{{\rm P}(3)} - z_{{\rm P}(4)})^2 } \Psi_{\rm L} \ ,
\end{align}
with ${\rm P}$ denoting permutations. Optimizing the wave function
$\Psi= \alpha \Psi_{\rm L} + \beta \Psi_{\rm trial}$ yields similar results as
the quasihole/quasiparticle constructions, that is, it captures well the
physics in the Laughlin-like regime, $V_2/V_0\lesssim 0.5$, but fails at 
larger values of $V_2/V_0$.

For larger values of $V_2/V_0$, it seems to be natural that more 
and more second-order anticorrelations are broken to free angular 
momentum for the creation of fourth-order anticorrelations. A wave 
function which pairwise removes anticorrelations from the Laughlin 
state is the Haffnian wave function~\cite{haffnian}: 
\begin{align}
 \Psi_{\rm Hf} \sim \sum_{\rm P} \frac{1}{(z_{{\rm P}(1)} - z_{{\rm P}(2)})^2 
\cdots (z_{{\rm P}(N-1)} - z_{{\rm P}(N)})^2} \Psi_{\rm L} \ .
\end{align}
Similarly to the Pfaffian construction by G. Moore and N. Read~\cite{MR} 
with $p$-wave pairing, the Haffnian wave function describes the formation 
of $d$-wave pairing. The Haffnian frees $N$ quanta of angular momenta 
which now can be used to generate fourth order anticorrelations. A 
natural way for reintroducing angular momentum is by another Haffnian, 
${\rm Hf} (z_i-z_j)^2$. In this way we obtain
\begin{align}
\label{hfhf}
 \Psi_{\rm Hf/Hf} \sim & \left[\sum_{\rm P} 
(z_{{\rm P}(1)} - z_{{\rm P}(2)})^2 \cdots 
(z_{{\rm P}(N-1)} - z_{{\rm P}(N)})^2 \right] \times \nonumber \\ &
\left[
\sum_{\rm P} \frac{1}{(z_{{\rm P}(1)} - z_{{\rm P}(2)})^2 \cdots 
(z_{{\rm P}(N-1)} - z_{{\rm P}(N)})^2}\right] \Psi_{\rm L} \ .
\end{align}
Alternatively, one might consider a quasihole construction, that is, 
the wavefunction $\Psi_{\rm Hf}^{\rm qh} \sim O_{\rm qh}(\xi=0)\Psi_{\rm Hf}$, 
which yields a similar expression as $\Psi_{\rm Hf/Hf}$, with overlap
$\Big|\braket{\Psi_{\rm Hf/Hf}}{\Psi_{\rm Hf}^{\rm qh}} \Big| = 0.97$ for 
$N=6$. Both wave functions yield squared overlaps $>0.5$ with the exact 
ground state for $V_2/V_0\gtrsim2$, as shown in Fig.~\ref{overlaps}.

\subsubsection{Two-body correlation functions}

Many-body correlated states, e.g. the Laughlin, are known 
to exhibit their inner order in two-body correlation functions, 
despite the fact that no structure is seen in the one body 
density. Here we present two-body correlation functions, 
see Fig.~\ref{corrs}, for selected values of $V_2/V_0$ and $N=6$. 
The plots show the probability distribution of finding a particle 
at position $(x,y)$ when another particle has been fixed at 
$(x,y)=(2.5,0)$, a position of relatively high density for any 
value of $V_2/V_0$ (cf. Fig.~\ref{densprof}). 

\begin {figure}[t]
\includegraphics[width=0.45\textwidth, angle=-90]{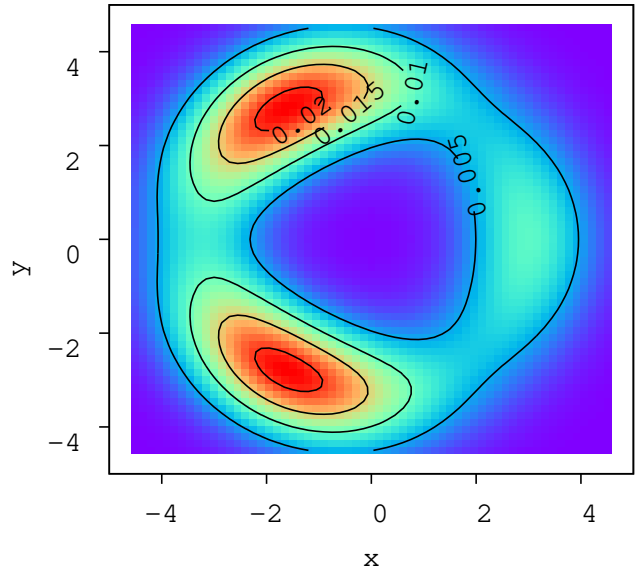}
\caption{(Color online)\label{fighaff} Pair correlation functions 
for $N=6$ for $\Psi_{\rm Hf/Hf}$ given in Eq.~(\ref{hfhf}) with one particle
fixed 
to $(x,y)=(2.5,0)$. }
\end {figure}

In the Laughlin state, Fig.~\ref{corrs}(a), the probability of 
finding two particles at the same position is zero. Apart from 
this strong anticorrelation, the state does not show pronounced 
correlation structures. Fixing one particle still allows the 
other particles to occupy the remaining area with some soft 
peaks near the edge of the system. The same peaks become more 
pronounced in Fig.~\ref{corrs}(b) at $V_2/V_0=0.3$, showing 
that 4 particles tend to arrange on a semicircle at the edge, 
while the fifth is most likely to be found near the center. 
The anticorrelation of the Laughlin is still present. The 
similarity between the correlation pattern in (a) and (b) 
suggest that both states describe the same phase. In (c), 
at $V_2/V_0=0.5$, the peaks of (b) get more pronounced, but 
the anticorrelation structure is lost. Increasing $V_2/V_0$ even 
more, to 0.8 in (d), 1.5 in (e), or 4.0 in (f), it becomes more 
and more likely to find a second particle at the position $(2.5,0)$. 
In fact, the structure in (d), with a peak at $(2.5,0)$ and two 
peaks on the opposite side of the cloud is a precursor of the 
pairing described by Eq.~(\ref{hfhf}). As is seen from 
Fig.~\ref{overlaps}, the overlap with this wave function starts 
to grow for $V_2/V_0>0.5$. Interestingly, the correlation structure 
of (d) does not transform continuously into the correlation 
structure of (f), with three very pronounced peaks, and almost zero 
values between the peaks. Instead, at intermediate values, as shown 
in (e) for $V_2/V_0=1.5$, there is only one peak at $(2.5,0)$ and 
increased values forming an semicircle on the other side of the cloud.
For $V_2/V_0=4$, the overlap with Eq.~(\ref{hfhf}) is $\simeq 0.8$, 
see Fig. \ref{overlaps}. Accordingly we find good resemblance
between the pair correlations in the ground state, Fig.~\ref{corrs}(f), 
and the pair correlation of the trial state, shown in Fig.~\ref{fighaff}, 
which clearly features a similar pairing structure.

Summarizing this discussion, the anti-correlated nature of the
Laughlin state is lost at $V_2/V_0\approx 0.5$, and the system
undergoes an evolution to a state with strong pairing and strong
anticorrelations between the pairs described well by the trial 
wave function Eq.~(\ref{hfhf}).

Let us also note that in the limit $V_2/V_0 \rightarrow \infty$, 
the pairing state is not the only ground state of the system. 
For instance, a state  $\Psi \sim Z^{N(N-1)}$ with $Z=\sum_i z_i$ is a 
zero-energy eigenstate of any potential with $V_0=0$,  
since all angular momentum is now simply attached to the center-of-mass. 
The number of other possibilities to achieve zero-energy states 
increases with $N$. This observation may seem to contradict the trend 
in the energy spectrum, Fig.~\ref{spectra}. In this figure, the 
increase of the gap for large $V_2/V_0$ is due to the choice of 
a constant $V_0$ defining the units of energy. Since the limit 
$V_2/V_0 \rightarrow \infty$, would then require 
$V_2 \rightarrow \infty$, the gap becomes small in comparison to 
the dominant energy scale of the system.

\section{Summary and discussion \label{outlook}}

In this paper we have studied few-body bosonic systems in the 
fractional quantum Hall regime, that is, in two dimensions, 
subjected to an external gauge field, and with repulsive atom-atom 
interactions. We have presented exact diagonalization results for small systems, 
with the aim of understanding how the topological properties of 
the many-body states, well understood in the thermodynamic limit 
for certain trial wave functions presented in Section~\ref{ideal}, 
manifest in few-body systems. 

Some of these ``trial''  states, notably the 1/2-Laughlin state for 
single component or the (221)-Halperin for 
two-component bosons, are actually exact ground states of 
the system with contact interactions. While in such a case it is a priori clear that the given state describes correctly the system of interest, it is not obvious how finite-size effects affect the topological features. Other trial states, for instance generic composite fermion states, have 
overlaps with the ground state of the system which decrease when the 
system size is increased. In this case, it is even harder to classify the topological order of the 
system by refering to some trial wave function. In this paper, by considering the fractionality of vortex 
excitations, that is the fractional effective charge 
of a single vortex, and the fractional statistics of two vortices, 
we have directly employed one of the most intriguing properties of 
topological order, and have thereby characterized different quantum Hall phases.

Our analysis in Section~\ref{calc} has shown that topological features are displayed already in very small systems. Both effective charge and 
braiding angle tend towards the expected bulk value in case of the 
1/2-Laughlin state, 1/4-Laughlin state, (221)-Halperin state for systems as small as 
 $N = 6$ particles. The most pronounced bulk is obtained for 
the 1/2-Laughlin state. Furthermore, we were able to use our study of vortex properties to distinguish between a state describing 
a rotationally symmetric quasiparticle excitation of the 1/2-Laughlin state from the composite 
fermion state at the same angular momentum $L=N(N-2)$. More precisely, we find that the effective 
charge of a vortex yields a good estimate of the filling factor, 
while no bulk is found with respect to the braiding of vortices. We 
note in this context that the quasihole charge, despite being related 
to the local density of the state, is not simply proportional to 
it. Thus, although the filling factor may already be estimated by 
a simple measurement of the density, such measurement is \textit{not} 
equivalent to the measurement of the effective charge.

In Section~\ref{perturbed} we considered a sytem described by 
the parent Hamiltonian of the Laughlin state plus a perturbation in form of $d$-wave scattering. Fixing the angular momentum to $L=N(N-1)$, we have studied the evolution of the system when increasing the perturbation. We could show that the braiding of vortices yields a signal of the topological phase transition occuring in the system, seen as an increased sensibility of the braiding angle on the positions of the vortices. 
Our estimate of the phase boundary obtained by the braiding is in agreement with estimates based on different quantities like the overlap with 
the Laughlin or Laughlin-like state, number of the edge excitations, and the energy spectrum at constant $L$. As the spectrum lacks ground state level crossings, we conclude that the topological phase transition occurs in a continuous way. In the regime where $d$-wave scattering is strong, we are able to  describe the system by a $d-$wave paired state, which is constructed from the Laughlin state by applying two different forms of Haffnian terms. This trial state has increasingly large overlap and exhibits similar pairing features as observed in the two-body correlation function of the exact ground state.

Demonstrating how topological order reflects in anyonic properties of small-sized systems, our study provides information which can be 
useful for validating a quantum simulation of fractional quantum Hall systems. In particular, such experiments might allow for variable 
system sizes, ranging from the sizes considered here, accessible through classical computations, to large numbers outperforming 
any classical attempt. As the comparison of different sizes from 4 to 7 shows, one can expect rapid improvements of the results only 
by a slight increase of the system size. We thus conclude that the measurement of the effective charge and the braiding angle, e.g. by 
using the scheme provided in Ref.~\cite{par2}, is a powerful tool for classifying the topological order of a system. In particular, such experiments could for the first time observe fractional quantum statistics in a direct way.

\begin{acknowledgements}
The authors thank N. Barber\'an for providing to us her manuscript on Haldane pseudopotentials. This work has been supported 
by EU (SIQS, EQUAM), ERC (QUAGATUA), Spanish MINCIN (FIS2008-00784
TOQATA), Generalitat de Catalunya (2009-SGR1289), and Alexander von Humboldt
Stiftung. BJD is supported by the Ram\'on y Cajal program.
\end{acknowledgements}

% \clearpage
% \bibliographystyle{apsrev_mod}
% \bibliography{bib.bib}

\end{document}